\documentclass[a4paper,12pt]{article}
\usepackage{epsfig}

\newcommand{\be}{\begin{equation}}
\newcommand{\ee}{\end{equation}}
\newcommand{\ba}{\begin{eqnarray}}
\newcommand{\ea}{\end{eqnarray}}
\newcommand{\ban}{\begin{eqnarray*}}
\newcommand{\ean}{\end{eqnarray*}}

\begin{document}

\title{\Large\bf Does the quantum collapse make sense?\\
Quantum Mechanics vs Multisimultaneity\\
in interferometer-series experiments}

\author{{\bf Antoine Suarez}\thanks{suarez@leman.ch}\\ Center for Quantum
Philosophy\\ The Institute for Interdisciplinary Studies\\ P.O. Box
304, CH-8044 Zurich, Switzerland}


\maketitle

\vspace{1cm}

\vspace{2cm}
\begin{abstract}

It is argued that the three assumptions of quantum collapse, `one
photon-one count', and relativity of simultaneity cannot hold
together: nonlocal correlations may depend on the referential
frames of the beam-splitters but not of the detectors. New
experiments using interferometers in series are proposed which make
it possible to test Quantum Mechanics vs Multisimultaneity.\\

{\em Keywords:} multisimultaneity, relativistic nonlocal causality,
wavefunction collapse, superposition principle.

\end{abstract}

\pagebreak

\section{Introduction}

Multisimultaneity (or Relativistic Nonlocality) is a description of
physical causality which unifies the relativity of simultaneity and
superluminal nonlocality avoiding superluminal signaling. The
description accounts in particular for the superluminal nonlocal
influences, and the consequent violation of relativistic causality,
happening in Bell experiments with space-like separated measuring
devices. Multisimultaneity deviates from Quantum Mechanics in that
two-particle correlations are supposed to depend on the timing of
the arrivals of the particles at the measuring devices. Whereas
both theories agree for all experiments conducted so far, they
conflict with each other in their predictions regarding new
proposed experiments with measuring devices in motion
\cite{asvs97.1, as97.2, as97.1, asvs97.2}.\\

Work to perform such experiments is in progress \cite{tbzg98.1,
tbzg98.2}. To test for Multisimultaneity it is crucial to set in
motion precisely those objects where take place the events which
are connected by superluminal influences. The already published
work \cite{asvs97.1, as97.2, as97.1} assumes that these events are
the arrivals of photons at beam-splitters, and, consequently, what
determines the timing is the state of motion and the position of
the beam-splitters. As pointed out in \cite{asvs97.1, as97.2} there
is experimental evidence against the hypothesis that the choice of
the output port by which one photon leaves a beam splitter depends
on which detector the other photon reaches. However, a version of
Multisimultaneity assuming nonlocal causal links between the
detectors seems also to be possible in principle, and would have
the advantage of keeping the key role that standard Quantum
Mechanics attributes to detection.\\

In this paper we describe an experiment involving pairs of
entangled photons running through a series of interferometers
(Fig.1). It is shown that a theory assuming both the `one
photon-one count' principle (the detection of only one photon
cannot produce more than one count) and the relativity of
simultaneity has to give up the quantum collapse, and cannot invoke
frame-dependent links between detections to explain the
correlations. The alternative explanation by means of influences
between the beam-splitters allow us within Multisimultaneity to
propose new interesting experiments with devices in motion. But it
also implies the possibility of producing so called two {\em
non-before} impacts \cite{as97.2}, one at each arm of the setup,
with beam-splitters at rest. It is argued that such an experiment
may allow us to decide between Multisimultaneity and timing
insensitive theories such as Quantum Mechanics without having to
set devices in motion. It is also highlighted that the inadequacy
of causal links between detections challenges the concept of
backward causation.\\

\begin{figure}[t]
\centering\epsfig{figure=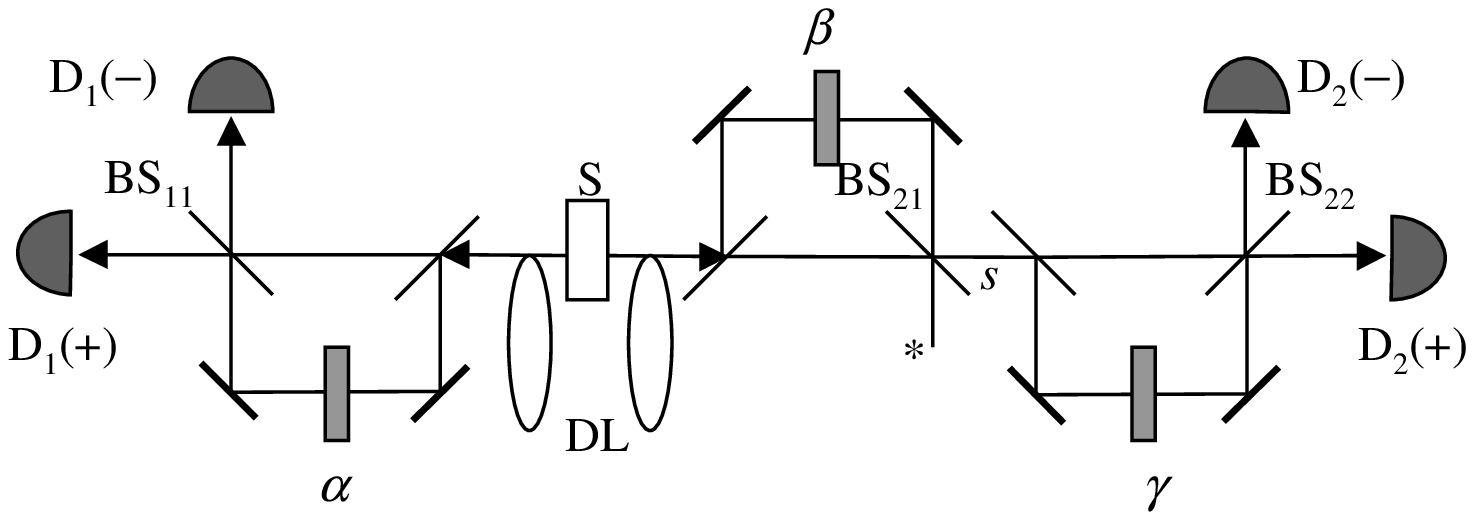,width=120mm}
{\small{\it{\caption{Impact series experiment with photon pairs:
photon 2 enters a two interferometer series impacting successively
on beam-splitter BS$_{21}$ and BS$_{22}$. See text for detailed
description.}}}}
\label{fig:BITfig1}
\end{figure}

\section{The experiment}

Consider the setup sketched in Fig.1. Energy-time entangled photon
pairs are emitted from a pulsed source S of the type described in
\cite{bgt98}. Photon 1 enters the left interferometer and impacts
successively on two beam-splitters before being detected in either
D$_{1}(+)$ or D$_{1}(-)$ after leaving beam-splitter BS$_{11}$.
Photon 2 enters a first interferometer on the right: if it leaves
BS$_{21}$ by output port $s$ it enters a second interferometer and
gets detected in either D$_{2}(+)$ or D$_{2}(-)$ after impacting
beam-splitter BS$_{22}$. Each interferometer, also the preparation
one within the source \cite{bgt98} (not sketched in Fig. 1),
consists in a long arm of length $L$, and a short one of length
$l$. We assume the path difference set to a value which largely
exceeds the coherence length of the photon pair light, and all
beam-splitters to be 50-50 ones.\\

We label the path pairs leading to detection as follows: $(l,lsl)$;
$(L,lsl)$; $(l,Lsl)$ and so on; where, e.g., $(l,Lsl)$ indicates
the path pair in which photon 1 has taken the short arm, and photon
2 has taken first the long arm, then the short one. We distribute
the ensemble of the 8 possible paths in the four following
subensembles:\\

\be
\begin{array}{lll}
(l,LsL)&:&2L-l\\
(L,LsL)\,,\,(l,Lsl)\,,\,(l,lsL)\,&:&L\\
(l,lsl)\,,\,(L,Lsl)\,,\,(L,lsL)\,&:&l\\
(L,lsl)&:&2l-L
\end{array}
\label{eq:paths}
\ee

where the right-hand side of the table indicates the relevant
parameter characterizing each subensemble of paths.\\

Time-resolved detection \cite{jb92,tbg97,ptjr94} of the photon
pairs cannot distinguish between the paths of subensemble $L$
because all of them will exhibit the same time difference in the
detected photon pair signals. Neither can measurement of the time
of emission of the pump laser light distinguish between these paths
when, as assumed, the path difference $L-l$ between the long and
the short path is the same for each interferometer included the
preparation one \cite{bgt98}. Therefore according to quantum
mechanics the paths of subensemble $L$ will interfere with each
other, and the same holds for the paths of subensemble $l$.\\

On the contrary, time-resolved detection allows us to discriminate
between paths of different subensembles in table \ref{eq:paths},
and in particular between the cases where a pair follows a path of
subensemble $L$, and the cases where the pair follows a path of
subensemble $l$. A time delay spectrum of coincidence counts
\cite{jb92,ptjr94} for each of the four possible outcomes
D$_{1}(\sigma)$, D$_{2}(\omega)$ \,($\sigma,\omega\in\{+,-\}$) will
exhibit four peaks: an interference peak corresponding to
subensemble $L$ we suppose set at time difference 0, a second
interference peak at time difference $(l-L)/c$ corresponding to
subensemble $l$, and two other peaks at time differences $(L-l)/c$
and $(2l-2L)/c$ corresponding to subensemble $2L-l$, respectively
$2l-L$. Using a time difference window one can select only the
events corresponding to subensemble $L$, or only those to
subensemble $l$.\\

For the sake of simplicity we refer to the different subpopulations
of detected photon pairs as subpopulation $L$, $l$ etc. Unless
stated otherwise, the experiments considered in the following are
supposed to involve only pairs of subpopulation $L$.\\

By means of delay lines DL different timings can be arranged in the
laboratory frame. The beam-splitters can also be supposed as in
motion, like in the experiments proposed in \cite{asvs97.1,
as97.2}.\\

\section{Timing insensitive Quantum Mechanics}

The quantum mechanical superposition principle states independently
of any possible timing:

\ba
P^{QM}(\sigma,\omega)&=&
\left|A(L\,\sigma,LsL\,\omega)+A(l\,\sigma,Lsl\,\omega)+
A(l\,\sigma,lsL\,\omega)\right|^2
\label{eq:JPQML}
\ea

where $P^{QM}(\sigma,\omega)$ denotes the joint probability of
getting the outcome D$_{1}(\sigma)$, D$_{2}(\omega)$ in experiments
with pairs of subpopulation $L$, and $A(path\,\sigma,path\,\omega)$
the corresponding probability amplitudes for the path and outcome
pairs specified within the parentheses.\\

Substituting the amplitudes into Eq. (\ref{eq:JPQML}) yields the
following values for the conventional joint probabilities:

\ba
P^{QM}(\sigma,\omega)
=\frac{1}{64}
\Big[3+\sigma\,2\cos(\alpha+\beta)+2\sigma\omega\,\cos(\alpha+\gamma)+\omega\,2\cos(\gamma-\beta)\Big]
\label{eq:qmjp}
\ea

\bigskip

From (\ref{eq:qmjp}) one is led to the following usual correlation
coefficient:

\ba
E^{QM}_{\sigma\omega}
=\frac{\sum_{\sigma,\,\omega}\sigma\omega\,P^{QM}(\sigma,\omega)}
{\sum_{\sigma,\,\omega}P^{QM}(\sigma,\omega)}
=\frac{2}{3}\cos(\alpha+\gamma)
\label{eq:qmccu}
\ea

and the special ones:

\ba
E^{QM}_{\sigma}
=\frac{\sum_{\sigma,\,\omega}\sigma\,P^{QM}(\sigma,\omega)}
{\sum_{\sigma,\,\omega}P^{QM}(\sigma,\omega)}
=\frac{2}{3}\cos(\alpha+\beta)
\label{eq:qmccsigma}
\ea

\ba
E^{QM}_{\omega}
=\frac{\sum_{\sigma,\,\omega}\omega\,P^{QM}(\sigma,\omega)}
{\sum_{\sigma,\,\omega}P^{QM}(\sigma,\omega)}
=\frac{2}{3}\cos(\gamma-\beta)
\label{eq:qmccomega}
\ea

Moreover, from Eq. (\ref{eq:qmjp}) one is led to the relation:\\

\ba
P^{QM}(\sigma,\pm)=\sum_{\omega}P^{QM}(\sigma,\omega)
=\frac{1}{32}\Big[3+\sigma\,2\cos(\alpha+\beta)\Big]
\label{eq:PsigmapmL1}
\ea

where $P^{QM}(\sigma,\,\pm)$ denotes the probability of getting
photon 1 detected in D$_{1}(\sigma)$ independently of where photon
2 is detected, for experiments with pairs of subpopulation $L$, or
in other words, the probability of getting a pair reaching
BS$_{11}$ and BS$_{22}$ by a path of subpopulation $L$, and photon
1 detected in D$_{1}(\sigma)$.

\bigskip

\section{Sharp defined relativistic nonlocal causal links}

As far as one looks upon correlated events as revealing some kind
of influence at work (``correlations cry out for explanation,''
John Bell said), three kinds of sharp defined frame-dependent
causal links can be considered candidates to explain the
correlations:

\begin{enumerate}
\item{{\em Detection-detection}: The choice of the detector into which photon $i$ falls
influences the choice of the detector into which photon $j$ falls.}
\item{{\em Detection-splitter}: The detection of photon $i$ influences the choice of
the output port by which photon $j$ leaves a beam-splitter.}
\item{{\em Splitter-splitter}: The choice of the output port by which photon $i$ leaves the
beam-splitter influences the choice of the output port by which
photon $j$ leaves a beam-splitter.}
\end{enumerate}

\bigskip

\subsection{Relativistic nonlocal influences between detections contradict
the `one photon-one count' principle}

\begin{figure}[t]
\centering\epsfig{figure=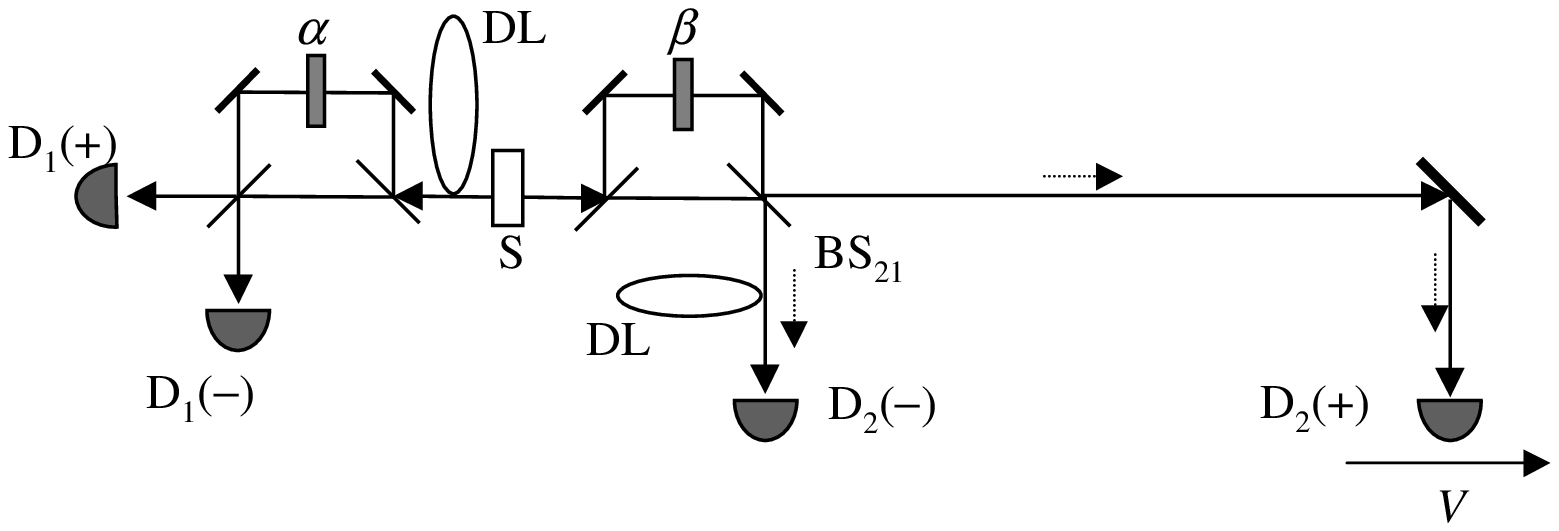,width=120mm}
{\small{\it{\caption{Experiment with detectors D$_{2}(\omega)$ in
relative motion to each other. See text for detailed
description.}}}}
\label{fig:BITfig1}
\end{figure}

Consider the simplified experiment sketched in Fig. 2 in which each
photon runs through only one interferometer, and assume that the
detection of photon 2 occurs time-like separated before the
detection of photon 1. The hypothesis that till the instant of
detection it is not determined which of the two detectors
D$_{2}(\omega)$ fires necessarily implies some kind of superluminal
influence or Bell connection between these detectors. Suppose that
D$_{2}(+)$ and D$_{2}(-)$ are set in relative motion to each other
so that the arrival of photon 2 at any D$_{2}(\omega)$, in the
inertial frame of D$_{2}(\omega)$, occurs before the arrival of
photon 2 to D$_{2}(-\omega)$. Under these conditions, the fact that
D$_{2}(\omega)$ fires, cannot depend on whether D$_{2}(-\omega)$
fires or not, and therefore it should happen that $25\%$ of the
times D$_{2}(+)$ and D$_{2}(-)$ fire together even if there is only
one particle traveling the right side of the setup, and $25\%$ of
the times neither D$_{2}(+)$ nor D$_{2}(-)$ fires when photon 2
reaches these detectors. This clearly contradicts the basic `one
photon-one count' principle. Therefore, as far as one keeps this
principle and the relativity of simultaneity, one has to reject
that the outcomes are determined by relativistic nonlocal
connections between D$_{2}(+)$ and D$_{2}(-)$, and accept they are
determined when the particles leave the beam-splitters. Notice that
this conclusion also holds for single particle experiments.
Moreover, the conclusion obviously implies that each particle
consists of a detectable part leaving the splitter by one of the
output ports, and an undetectable part leaving by the other. \\

In summary, realtivistic nonlocal causal links between detectors
seem inadequate to explain the correlations.\\

\subsection{Experiment rules out detection-splitter links}

Even if the outcomes are determined when the particles leave the
beam-splitters, as stated in Subsection 4.1, one could nevertheless
imagine that the correlations appear because the detection of
particle $j$ affects the outcome choice of particle $i$ at the
corresponding beam-splitter.\\

Suppose a conventional Bell-experiment with detectors set at
distances such that the detection of each particle lies always
time-like separated after the arrival of the other particle at the
beam-splitter. As referred to in \cite{asvs97.1}, the experiment
used in \cite{jrpt90} fulfills this condition.\\

For such an experiment, the considered detection-splitter
hypothesis would imply the disappearance of the correlations, for
none of the particles can be affected by any influence from the
other side. The experimental results in  \cite{jrpt90} contradict
this prediction, and rule out the hypothesis that the correlations
arise because detection-splitter links.\\

\subsection{Splitter-splitter links, the remaining explanation}

In conclusion, any causal explanation sharing the `one photon-one
count' principle and the relativity of simultaneity has to assume
that what matters for the appearance of correlations are
frame-dependent links between events at the beam-splitters.
Detection only reveals choices which have taken place already, and
does not play any particular role in determining which outcome
occurs.\\

\section{Timing-dependent Multisimultaneity}

In agreement with the conclusions of Section 4 Multisimultaneity
assumes that the outcome of an experiment depends on the relative
state of motion and position of the beam-splitters in which
interferences take place, and not of the detectors, for instance,
in the experiment of Fig.1, the beam-splitters BS$_{11}$,
BS$_{21}$, and BS$_{22}$. Then, arbitrary large series of
interferometers make it possible to arrange plenty of different
timings, beyond the three already discussed in the realm of
conventional Bell setups, i.e., two {\em before}, one {\em
non-before}, and two {\em non-before} Timings \cite{asvs97.1,
as97.2}, and obviously Multisimultaneity has to account for all of
them.\\

\subsection{Principles of Multisimultaneity}

First of all we generalize the main concepts and principles in
\cite{asvs97.1, as97.2} in order to describe experiments with
series of interferometers.\\

We assume that each particle $i$ consists of an observable or
detectable part traveling by one of the paths, and several
unobservable parts traveling by the other possible paths at the
same speed as the observable one. Moreover we assume that the
impacts of the observable parts of the two photons on the
beam-splitters are connected by means of superluminal influences.\\

We denote $(T_{jl})_{ik}$  ($i,j \in\{1,2\}; k,l
\in\{1,2...n\}$) the time at which the observable part of particle $j$
impacts on beam-splitter BS$_{jl}$, measured in the inertial frame
of BS$_{ik}$ (the subscript $ik$ after a parenthesis means all
times within the parentheses to be measured in the inertial frame
of BS$_{ik}$).\\

At time $(T_{ik})_{ik}$, we consider the latest BS$_{jl}$, if any,
such that $(T_{jl}\leq T_{ik})_{ik}$. If at any time $T$ such that
$(T_{ik}<T<T_{i\,k+1})_{ik}$ and $(T_{jl}<T<T_{j\,l+1})_{ik}$, it
is impossible to distinguish by which path the particles did travel
before leaving the beam-splitters BS$_{ik}$ and BS$_{jl}$, the
impact on BS$_{ik}$ is said to be {\em non-before} the impact on
BS$_{jl}$, and denoted $a_{ik[jl]}$, or simply $a_{ik}$ if no
ambiguity results. Otherwise, the impact on BS$_{ik}$ is called a
{\em before} one, and denoted $b_{ik}$. Expressions like
$a_{ik[jl]\,\sigma}$, $b_{ik\,\sigma}$ denote that the detectable
part of particle $i$ leaves BS$_{ik}$ by the output port $\sigma$
in the indicated {\em non-before}, respectively {\em before}impact.
One could say that a photon undergoing a {\em non-before} impact
can consider, from ``its point of view'', indistinguishability
guaranteed if detections occur, and a photon undergoing a {\em
before} impact cannot.\\

According to these definitions, the conventional Bell tests are
referred to as $(b_{i}, a_{j})$ experiments, for all beam-splitters
are at rest, and one of the impacts always occurs before the other
in the laboratory frame. The Timing considered in Section 4 is
referred to as $(b_{11}, a_{21}\, a_{22})$. Besides this Timing, we
will be interested in the following two other ones: $(b_{11},
b_{21}\, a_{22})$,  and $(a_{11[21]}, b_{21}\, a_{22})$.\\

Multisimultaneity rests on two main principles:

\begin{enumerate}
\item{{\em Principle I}: Values $b_{ik\,\sigma}$ do not depend on phase
parameters particle $j$ meets.}
\item{{\em Principle II}: Values $a_{ik[jl]\,\sigma}$ do not
depend on values $a_{jl'[ik']\,\omega'}$ ($l'\leq l$) particle $j$
may actually produce, but only of the values $b_{jl'\,\omega}$
particle $j$ would have produced in corresponding {\em before}
impacts.}
\end{enumerate}

Because of the Bell experiments already conducted, {\em Principle
II} obviously implies that in experiments $(b_{i}, a_{j})$ the
distribution of the outcome results is calculated combining the
amplitudes of the single alternative paths, the same way as Quantum
Mechanics does. This rule is extended to experiments in which all
the impacts of particle $i$ are {\em before}, and all those of
particle $j$ are {\em non-before}, i.e., in such experiments the
conventional Quantum Mechanical rule of combining amplitudes
applies. Regarding experiments like $(b_{11}, b_{21}\, a_{22})$ in
which particle $i$ undergoes only {\em before} impacts, and
particle $j$ undergoes some {\em before} and some {\em non-before}
ones, the outcome's distribution is still calculated combining
amplitudes, but according rules unknown in Quantum Mechanics, as we
will see in Subsection 5.3.\\

In experiments involving an $a_{ik[jl]}$ impact and an
$a_{jl'[ik]}$ one, it would be absurd to assume together that the
impacts on BS$_{ik}$ take into account the actual outcomes of the
impacts on BS$_{jl'}$, and that the impacts on BS$_{jl'}$ take into
account the actual outcomes of the impacts on BS$_{ik}$. Therefore,
Timing $(a_{11[22]}, b_{21}a_{22})$ requires that
$a_{11[22]\,\sigma}$ depends on $b_{21}b_{22\,\omega'}$, and
$b_{21}a_{22\,\omega}$ on $b_{11\,\sigma'}$; Timing $(a_{11[22]},
a_{21} a_{22})$ that $a_{11\,\sigma}$ depends either on
$b_{21\,\ast}$, or on $b_{21\,s\,l}$, or on
$b_{21}b_{22\,\omega'}$, and $a_{21}a_{22\,\omega}$ depends on
$b_{11\,\sigma'}$. The motivation of {\em Principle II} is to
account for all Timings involving {\em non-before} impacts without
multiplying hypothesis needlessly.\\

Assuming the relativity of simultaneity and excluding backward
causation, Multisimultaneity consequently forbids superluminal
signaling. This impose constraints to the path amplitudes. It
implies in particular that for any non-selective experiment the
probability that particle $i$ produces the result $\sigma$
(independently of which outcome particle $j$ produces) does not
depend on timing.\\

\subsection{Timing $(b_{11}, a_{21}\, a_{22})$}

Such a Timing can be easily arranged by keeping all beam-splitters
at rest, at distances such that the impact on BS$_{11}$ occurs
before the impact on BS$_{21}$.\\

Regarding the pairs whose detectable parts travel path $(l,L)$,
photon 2 at its arrival at BS$_{21}$ cannot consider
indistinguishability guaranteed, and therefore the probability of
getting photon 1 detected in D$_{1}(\sigma)$, and the detectable
part of photon 2 leaving BS$_{21}$ by output port $s$ is given by
the relation:

\ba
P(b_{11\,\sigma}, b_{21\, s})
=P(l,L)P(b_{11\,\sigma}|l,L)P(b_{21\,s}|l,L)
\label{eq:Pb11sigmab21s}
\ea

where $P(l,L)$ means the probability of having a pair with
detectable parts traveling path $(l,L)$, $P(b_{11\,\sigma}|l,L)$
the probability that photon 1's detectable part of such a pair
leaves BS$_{11}$ by output port $\sigma$, and $P(b_{21\,s}|l,L)$
the probability that photon 2's detectable part leaves BS$_{21}$ by
output port $s$. Therefore it holds that:\\

\ba
P(b_{11\,\sigma}, b_{21\, s})
=\frac{1}{2^{2}}\frac{1}{2}\frac{1}{2}
=\Big|A(l\,\sigma,Ls)\Big|^2
=\frac{1}{2^{4}}
\label{eq:Pb11sigmab21s2}
\ea

\bigskip

Consider now the pairs whose detectable parts travel one of the
paths $(L,L)$ or $(l,l)$. Since photon 2 at its arrival at
BS$_{21}$ can consider indistinguishability guaranteed, the
probability of getting photon 1 detected in D$_{1}(\sigma)$, and
the detectable part of photon 2 leaving BS$_{21}$ by output port
$s$ is given by the relation:

\ba
P(b_{11\,\sigma}, a_{21\, s})
&=&P(l,l)P(b_{11\,\sigma}|l,l)P(a_{21\,s}|b_{11\,\sigma})\nonumber\\
&+&P(L,L)P(b_{11\,\sigma}|L,L)P(a_{21\,s}|b_{11\,\sigma})
\label{eq:Pb11sigmaa21s}
\ea

where $P(a_{21\,s}|b_{11\,\sigma})$ denotes the probability that in
a pair undergoing impacts $(b_{11}, a_{21})$ photon 2's observable
part leaves BS$_{21}$ by output port $s$, assuming photon 1's
observable part did leave BS$_{11}$ by output port $\sigma$. We
assume this conditional probability to be given by the relation:

\ba
P(a_{21\,s}|b_{11\,\sigma})
&=&\frac{|A(L\,\sigma,L\,s)+A(l\,\sigma,l\,s)|^2}
{\sum_{\omega}|A(L\,\sigma,L\,\omega)+A(l\,\sigma,l\,\omega)|^2}\vspace{0.2cm}\nonumber\\
&=&\frac{1}{2}[1+\sigma \, \cos(\alpha+\beta)]
\label{eq:Pa21|b11}
\ea

Substituting (\ref{eq:Pa21|b11}) into (\ref{eq:Pb11sigmaa21s}) one
gets:

\ba
P(b_{11\,\sigma}, a_{21\, s})
&=&\Big|A(L\,\sigma,Ls)+A(l\,\sigma,ls)\Big|^2\nonumber\\
&=&\frac{1}{2^{4}}\Big[2+\sigma \,2\cos(\alpha+\beta)\Big]
\label{eq:Pb11sigmaa21s2}
\ea

From Eq. (\ref{eq:Pb11sigmaa21s2}) and (\ref{eq:Pb11sigmab21s2}) it
follows that the probability of a pair reaching BS$_{11}$ and
BS$_{22}$ by a path of subpopulation $L$, and producing outcome
D$_{1}(\sigma)$ is given by the relation:

\ba
P(b_{11\,\sigma}, a_{21}\, a_{22\,\pm})
&=&\frac{1}{2}\Big|A(L\,\sigma,Ls)+A(l\,\sigma,ls)\Big|^2
+\frac{1}{2}\Big|A(l\,\sigma,Ls)\Big|^2\nonumber\\
&=&\Big|A(L\,\sigma,LsL)+A(l\,\sigma,lsL)\Big|^2
+\Big|A(l\,\sigma,Lsl)\Big|^2\nonumber\\
&=&\frac{1}{2^{5}}\Big[2+\sigma\,2\cos(\alpha+\beta)\Big]+\frac{1}{2^{5}}
\label{eq:PsigmapmL2}
\ea

\bigskip

which agrees with the prediction (\ref{eq:PsigmapmL1}) of Quantum
Mechanics.

\bigskip

Consider the pairs reaching BS$_{11}$ and BS$_{22}$ by a path of
subpopulation $L$, and photon 1 yielding outcome value $\sigma$.
Photon 2 of these pairs at its arrival at BS$_{22}$ can consider
indistinguishability guaranteed. Assume the probability that photon
2 of such a pair yields outcome value $\omega$ to be given by the
expression:

\ba
P(a_{21}a_{22\,\omega}|b_{11\,\sigma})
=\frac{\Big|A(L\,\sigma,LsL\,\omega)+A(l\,\sigma,Lsl\,\omega)+
A(l\,\sigma,lsL\,\omega)\Big|^2}
{\sum_{\omega}\Big|A(L\,\sigma,LsL\,\omega)+A(l\,\sigma,Lsl\,\omega)+
A(l\,\sigma,lsL\,\omega)\Big|^2}\vspace{0.2cm}
\label{eq:cpto2}
\ea

From Eq. (\ref{eq:PsigmapmL2}) and (\ref{eq:cpto2}) one is led to:

\ba
&&P(b_{11\,\sigma}, a_{21} a_{22\,\omega})\nonumber\\
&&\hspace{0.5cm}=P(a_{21}a_{22\,\omega}|b_{11\,\sigma})\,
\Big[\,|A(L\,\sigma,LsL)+A(l\,\sigma,lsL)|^2
+|A(l\,\sigma,Lsl)| ^2\Big]\nonumber\\
&&\hspace{0.5cm}=P^{QM}(\sigma,\,\omega)
\label{eq:t2}
\ea

This means that the quantum mechanical predictions can be explained
straightforwardly in a causal way, and the ``causal
indistinguishability condition'' proposed in \cite{as98.1} becomes
superfluous.\\

\subsection{Timing $(b_{11}, b_{21}\, a_{22})$}

Experiments holding this Timing could for instance be arranged by
setting BS$_{11}$ in motion so that $(T_{11}<T_{21})_{11}$, and
keeping BS$_{21}$ and BS$_{22}$ at rest at distances such that:
$T_{21}<T_{11}$ and $T_{11}<T_{22}$, these times measured in the
laboratory frame.\\

Since now photon 2 at its arrival at BS$_{21}$ cannot consider
indistinguishability guaranteed, the probability of getting a pair
reaching  BS$_{11} and $ BS$_{22}$ by a path of subpopulation $L$,
and photon 1 detected in D$_{1}(\sigma)$ is given by the relation:

\ba
P(b_{11\,\sigma}, a_{22\, \pm})
&=&P(L,L)P(b_{11\,\sigma}|L,L)P(b_{21\,s\,L}|L,L)\nonumber\\
&+&P(l,l)P(b_{11\,\sigma}|l,l)P(b_{21\,s\,L}|l,l)\nonumber\\
&+&P(l,L)P(b_{11\,\sigma}|l,L)P(b_{21\,s\,l}|l,L)
\label{eq:Pb11sigmaa22pm}
\ea

where $P(L,L)$ means the probability of having a pair with
detectable parts traveling path $(L,L)$, $P(b_{11\,\sigma}|L,L)$
the probability that photon 1's detectable part of such a pair
leaves BS$_{11}$ by output port $\sigma$, $P(b_{21\,s\,L}|L,L)$ the
probability that photon 2's detectable part of such a pair leaves
BS$_{21}$ by output port $s$ and thereafter enters the second
interferometer by the long arm $L$, and so on for the other
terms.\\

Therefore it holds that:

\ba
P(b_{11\,\sigma}, a_{22\, \pm})
=\frac{3}{2^{5}}
\label{eq:Pb11sigmaa22pm2}
\ea

which clearly contradicts the Quantum Mechanical prediction of Eq.
(\ref{eq:PsigmapmL1}).\\

Therefore, if one accepts {\em Principle I}, one cannot accept for
the interferometer-series experiment we are considering that the
outcome results are distributed according to the Quantum Mechanical
superposition principle.\\

To account for the new situation Multisimultaneity assumes the
following rule which is unknown in Quantum Mechanics:

\ba
P(b_{11\,\sigma},b_{21}\, a_{22\omega})
=\Big|A(L\,\sigma,LsL\,\omega)\Big|^2
+\Big|A(l\,\sigma,Lsl\,\omega)+A(l\,\sigma,lsL\,\omega)\Big|^2\nonumber\\
+A(L\,\sigma,LsL\,\omega)A^{*}(l\,\sigma,Lsl\,\omega)
+A^{*}(L\,\sigma,LsL\,\omega)A(l\,\sigma,Lsl\,\omega)\nonumber\\
=\frac{1}{64}
\Big[3+\sigma\omega 2\cos(\alpha+\gamma)+\omega 2\cos(\gamma-\beta)\Big]
\label{eq:Pb11a21a22}
\ea

This rule can easily be generalized to all possible timings arising
in experiments with arbitrary large series of interferometers as
shown in another article.\\

From Eq. (\ref{eq:Pb11a21a22}) one gets the following conditional
probabilities:

\ba
P(b_{21} a_{22\,\omega}|b_{11\,\sigma})
&=&\frac{P(b_{11\,\sigma},b_{21} a_{22\,\omega})}
{P(b_{11\,\sigma},b_{21} a_{22\,\pm})}\nonumber\\ &=&\frac{1}{6}
\Big[3+\sigma\omega \,2\cos(\alpha+\gamma)+\omega \,2\cos(\gamma-\beta)\Big]
\label{eq:Pb21a22|b11}
\ea

where $P(b_{21}a_{22\,\omega}|b_{11\,\sigma})$ denotes the
probability that in a pair undergoing impacts $(b_{11}, b_{21}
a_{22})$ photon 2's observable part leaves BS$_{22}$ by output port
$\omega$, assuming photon 1's observable part did leave BS$_{11}$
by output port $\sigma$.\\

Eq. (\ref{eq:Pb11a21a22}) yields the usual correlation coefficient:

\ba
E_{\sigma\omega}
=\frac{\sum_{\sigma,\,\omega}\sigma\,\omega\,P(\sigma,\omega)}
{\sum_{\sigma,\,\omega}P(\sigma,\,\omega)}
=\frac{2}{3}\cos(\alpha+\gamma)
\label{eq:MSccu1}
\ea

which agrees with the prediction (\ref{eq:qmccu}) of Quantum
Mechanics, the special $E_{\sigma}$ one:

\ba
E_{\sigma}
=\frac{\sum_{\sigma,\,\omega}\sigma\,P(\sigma,\omega)}
{\sum_{\sigma,\,\omega}P(\sigma,\omega)}
=0
\label{eq:MSccsigma1}
\ea

which contradicts the prediction (\ref{eq:qmccsigma}) of Quantum
Mechanics, and the special $E_{\omega}$ one:\\

\ba
E_{\omega}
=\frac{\sum_{\sigma,\,\omega}\omega\,P(\sigma,\omega)}
{\sum_{\sigma,\,\omega}P(\sigma,\omega)}
=\frac{2}{3}\cos(\gamma-\beta)
\label{eq:MSccomega1}
\ea

which agrees with the prediction (\ref{eq:qmccomega}) of Quantum
Mechanics.\\

The conclusion that Timings $(b_{11}, b_{21}\, a_{22})$ and
$(b_{11}, a_{21}\, a_{22})$ do not yield the same joint
probabilities supersedes the version of Relativistic Nonlocality
discussed in \cite{as97.3}, whereas basically agrees with the
corresponding assumption of \cite{as97.1}.

\subsection{Timing $(a_{11[21]}, b_{21}\,
a_{22})$}

Such a Timing can be arranged by keeping all beam-splitters at
rest, at distances such that: $T_{11}<T_{21}<T_{22}$, all times
measured in the laboratory frame, i.e.: the impact on BS$_{21}$
occurs before the impact on BS$_{11}$, and the impact on BS$_{11}$
occurs before the impact on BS$_{22}$.\\

Consider the arrival of photon 1 at BS$_{11}$. For observable
particle parts traveling by path $(l, l)$ or $(L, L)$ one should
now assume that the output port by which photon 1 leaves BS$_{11}$
depends nonlocally on which output port photon 2 takes at
BS$_{21}$. Therefore it holds that:

\ba
P(a_{11[21]\,\sigma}|b_{21\,s})
&=&\frac{|A(L\,\sigma,L\,s)+A(l\,\sigma,l\,s)|^2}
{\sum_{\sigma}|A(L\,\sigma,L\,s)+A(l\,\sigma,l\,s)|^2}\vspace{0.2cm}\nonumber\\
&=&\frac{1}{2}[1+\sigma \, \cos(\alpha+\beta)]
\label{eq:Pa11[21]b21}
\ea

\bigskip

Consider now the arrival of photon 2 at BS$_{22}$. According to
{\em Principle II} photon 2 takes account of the value
$b_{11\,\sigma}$ photon 1 would have produced if it had arrived at
BS$_{11}$ before photon 2 arrived at BS$_{21}$. This yields the
following relation:

\ba
&&P(a_{11[21]\,\sigma},b_{21} a_{22\,\omega})\nonumber\\
&&\hspace{0.2cm}=\sum_{\sigma'}P(L,L)P(b_{11\,\sigma'}|L,L)P(b_{21\,s\,L}|L,L)
\,P(a_{11[21]\,\sigma}|b_{21\,s})\,P(b_{21}a_{22\,\omega}|b_{11\,\sigma'})\nonumber\\
&&\hspace{0.2cm}+\sum_{\sigma'}P(l,l)P(b_{11\,\sigma'}|l,l)P(b_{21\,s\,L}|l,l)
\,P(a_{11[21]\,\sigma}|b_{21\,s})\,P(b_{21}a_{22\,\omega}|b_{11\,\sigma'})\nonumber\\
&&\hspace{0.2cm}+\,P(l,L)P(b_{11\,\sigma}|l,L)P(b_{21\,s\,l}|l,L)\,P(b_{21}a_{22\,\omega}|b_{11\,\sigma})
\label{eq:Pa11[21]a22}
\ea

\bigskip

Substitutions into Eq. (\ref{eq:Pa11[21]a22}) according to Eq.
(\ref{eq:Pb21a22|b11}), and (\ref{eq:Pa11[21]b21}) lead to the
following joint probabilities:

\ba
&&P(a_{11[21]\,\sigma}, b_{21} a_{22\,\omega})
=\frac{1}{32}\frac{1}{6}
[3+\sigma\omega \,2\cos(\alpha+\gamma)+\omega
\,2\cos(\gamma-\beta)]\nonumber\\
&&\hspace{1cm}+\,\frac{1}{32}
\frac{1}{6}[1+\sigma \,\cos(\alpha+\beta)]
[6+\omega
\,4\cos(\gamma-\beta)]\nonumber\\
&&\hspace{1cm}=\frac{1}{32}\frac{1}{6}
\Big[9+\sigma \,6\cos(\alpha+\beta)
+\sigma\omega\,2\cos(\alpha+\gamma)\nonumber\\
&&\hspace{3.2cm}+\,\omega\,6\cos(\gamma-\beta)
+\sigma\omega\,4\cos(\alpha+\beta)\cos(\gamma-\beta)\Big]
\label{eq:Pmsjp}
\ea

\bigskip

And Eq. (\ref{eq:Pmsjp}) yields the following usual correlation
coefficient:

\ba
E_{\sigma\omega}=\frac {\sum_{\sigma,\,\omega}\sigma\,\omega
P(\sigma,\omega)} {\sum_{\sigma,\,\omega}P(\sigma,\omega)}
=\frac{2}{9}\Big[\cos(\alpha+\gamma)+2\cos(\alpha+\beta)\cos(\gamma-\beta)\Big]
\label{eq:MSccu2}
\ea

which differs from the prediction (\ref{eq:qmccu}) of Quantum
Mechanics, and the special ones:

\ba
E_{\sigma}
=\frac{\sum_{\sigma,\,\omega}\sigma\,P(\sigma,\omega)}
{\sum_{\sigma,\,\omega}P(\sigma,\omega)}
=\frac{2}{3}\cos(\alpha+\beta)
\label{eq:MSccsigma2}
\ea

\ba
E_{\omega}
=\frac{\sum_{\sigma,\,\omega}\omega\,P(\sigma,\omega)}
{\sum_{\sigma,\,\omega}P(\sigma,\omega)}
=\frac{2}{3}\cos(\gamma-\beta)
\label{eq:MSccomega2}
\ea

which agree with the predictions (\ref{eq:qmccsigma}) and
(\ref{eq:qmccomega}) of Quantum Mechanics.\\

Experiments with Timings $(a_{11[22]}, b_{21} a_{22})$,
$(a_{11[22]}, a_{21} a_{22})$, and $(a_{11[21]}, a_{21} a_{22})$
can basically be calculated the same way.\\

\section{Real experiments}

A first real experiment can be carried out adapting the setup
required to perform the experiments proposed in \cite{asvs97.1,
as97.2}: the photon impacting the beam-splitter at rest should be
led to enter a second interferometer before getting detected, and
the moving beam-splitter set so that $b_{11}$ and $b_{21}$ impacts
result. Then, for phase values:

\ba
\alpha =\beta =0^{\circ}
\label{eq:values1}
\ea

Eq. (\ref{eq:qmccsigma}) and (\ref{eq:MSccsigma1}) bear the
following contradictory predictions:

\ba
E^{QM}_{\sigma}&=&\frac{2}{3}\nonumber\\
E_{\sigma}&=&0
\label{eq:pred1}
\ea

\bigskip

A second real experiment without devices in motion can be carried
out arranging the conventional Bell setup used in \cite{tbg97}, in
order that one of the photons enters a second interferometer before
getting detected. For phase values:

\ba
\alpha =\gamma =0^{\circ}
\label{eq:values2}
\ea

Eq. (\ref{eq:qmccu}) and (\ref{eq:MSccu2}) bear the contradictory
predictions:

\ba
E^{QM}_{\sigma\omega}&=&\frac{2}{3}\nonumber\\
E_{\sigma\omega}&=&\frac{2}{9}(1+2\cos^2\beta)
\label{eq:pred2}
\ea

i.e.: Quantum Mechanics predicts a usual correlation coefficient
that does not depend on parameter $\beta$, whereas according to
Multisimultaneity the correlation coefficient should oscillate
between $\frac{2}{9}$ and $\frac{2}{3}$ as $\beta$ varies linearly
in time.\\

The experimental quantities corresponding to the different
correlation coefficients can be determined as usual through the
four measured coincidence counts $R_{\sigma\omega}$ in the
detectors. Work to realize these experiments is in progress.\\

\section{Some comparative remarks}

On the one hand Multisimultaneity shares the spirit of Quantum
Mechanics in that indistinguishability can still be considered a
sufficient condition for combining amplitudes to calculate the
outcome distribution. However within Multisimultaneity
indistinguishability is supposed to be established by observers in
different inertial frames, and according to the resulting variety
of experimental situations, different rules of combining amplitudes
may apply, instead of the only one used by Quantum Mechanics, the
superposition principle.\\

On the other hand, the spirit of Multisimultaneity looks somewhat
like the reverse of that animating Quantum Mechanics. The
impossibility of any kind of backward-in-time influences, and the
possibility of superluminal ones providing they cannot be used for
signaling, have in Multisimultaneity the status of principles. By
contrast, neither the impossibility of backward-in-time influences,
nor that of superluminal signaling are principles of Quantum
Mechanics, but result as theorems, i.e., as consequences of the
formalism which ``miraculously'' (since not aimed) permit  the
``pacific coexistence'' with Relativity.\\

As long as single-particle experiments are considered, the
Multisimultaneity description by means of observable and
unobservable particle's parts does not basically differ from Bohm's
``empty wave''. However, for multiparticle experiments both
descriptions clearly deviate: The ``quantum potential'' related to
the ``empty wave'', although acting in a superluminal way, is
supposed to carry only information regarding phase parameters,
remaining insensitive to relativistic timing. By contrast,
Multisimultaneity clearly establishes two different levels of
unobservable causes or ``veiled reality'' \cite{de95} (two classes
of ``empty waves'', one could say), stating that, firstly, the kind
of ``nonlocality'' that may be invoked to explain single-particle
interferences originates from subluminal influences and involves
only information about phase parameters, and, secondly, the
superluminal influences causing nonlocal multiparticle correlations
carry also information about the state of motion and the position
of the beam-splitters.\\

\section{Does the quantum collapse make sense?}

The analysis of Section 4, about which links can be invoked to
explain the correlations, may also stimulate us to reach a sharper
picture of what the ``wavefunction collapse'' may physically
mean.\\

On the one hand, if one rigorously ties ``collapse'' to detection,
and considers it the cause leading the system to jump into a
particular outcome value among several possible ones, then it
appears that the three principles of collapse, `one photon-one
count', and relativity of simultaneity cannot hold together.
Quantum collapse and `one photon-one count' imply quantum aether.\\

On the other hand, if one keeps the relativity of simultaneity
(because Michelson-Morley and related observations) and the basic
principle `one photon-one count', then one is obliged to assume
that detection does not play any role in determining which outcomes
an experiment produces but only matters to make irreversible the
decisions reached at the beam-splitters. But then the whole talk
about `collapse' and `superposition' seems to become superfluous.\\

Anyway, the analysis challenges the very concept of ``wavefunction
collapse''.

\section{Challenging Backward Causation}

The opposite view to the causal one is undoubtedly
``Retrocausation'', i.e., the position asserting that decisions at
present can influence the past.  ``Retrocausation'' has been
developed as a consistent Lorentz-invariant interpretation of
ordinary Quantum Mechanics by O. Costa de Beauregard \cite{co97}.
The discussion about the possibility of influences acting backward
in time has been recently stimulated by H. Stapp \cite{hs97}.\\

Regarding Retrocausation it is important to realize that
Multisimultaneity makes it possible to harmonize the causality
principle and superluminal nonlocality; therefore, speaking of
``backward-in-time influences'' makes sense only if such influences
can be demonstrated to exist between time-like separated regions
\cite{co98}. Only such a specific experiment may allow us to decide
between the causal view and retrocausation, similar to how Bell
experiments allow us to decide between local realism and
superluminal nonlocality.\\

Suppose that the interferometer-series experiment of Fig.1 were
realized according to the following Timing: The impact on BS$_{11}$
and detection at D$_{1}(\sigma)$ lie time-like separated before the
impact on BS$_{21}$. Could such an experiment be considered a
candidate to the aim of deciding between the causal view and
retrocausation?\\

This would be the case if, invoking Wheeler's Great Smoky Dragon
\cite{whee86}, one denies the right to speak about what is present
between the place where photon 2 enters the equipment at the first
half-silvered mirror and the place where it reaches one counter or
the other. For then no event on the left-hand side could be
supposed to determine which path photon 2 travels, and therefore
post-selection of subpopulations by time-resolved detection could
not make the probabilities for different detectable results on side
1 depend on a parameter set on the side 2 of the apparatus. Hence,
vindication of the single probabilities of Eq.
(\ref{eq:PsigmapmL1}) by the experiment would seem to reveal an
effect of the detection of photon 2 on the detection of photon 1.\\

However, if one accepts for the pairs of subpopulation $L$ that
photon 2 travels path segment $s$ before any detection occurs, it
is quite possible to explain things in a causal way by means of
splitter-splitter links, as shown in Subsection 5.2. So ``backward
causation'' would require that one cannot really say anything about
photon 2 between the instant it enters the first interferometer and
the instant it gets detected, not even that it enters the second
interferometer by path $s$ connecting the two interferometers on
the right. Undoubtedly this is hard to swallow, and the proposed
experiments also challenge the concept of retrocausation.\\

\section{Conclusion}

We have shown that frame-dependent links between detections is not
an adequate way to explain nonlocal correlations, and, more
specifically, that the option of setting detectors in motion to
test Quantum Mechanics means in fact to question the principle `one
photon-one count'.\\

Therefore, regarding Multisimultaneity one is led to conclude that
one of the particles chooses the output port by which to leave a
beam-splitter taking into account which choice the other particle
makes at the beam-splitters it meets. This conclusion allows us to
design new experiments to test Multisimultaneity vs Quantum
Mechanics, and in particular an experiment with one {\em
non-before} impact at each arm of the setup, without devices in
motion. Upholding of Multisimultaneity would demonstrate the
Relativistic Nonlocal Causal description to embrace more phenomena
than the Quantum Mechanical one. Rejection of Multisimultaneity in
the experiment without motion would rule out Principle II of the
theory.\\

Regarding Quantum Mechanical theories assuming timing-independent
correlations, the inadequacy of links between detections challenges
both, the collapse description and the attempt to save
Lorentz-invariance through influences backward-in-time. Upholding
of the quantum mechanical predictions by proposed experiments could
surely be interpreted in terms of theories assuming absolute
space-time, such as Bohm's theory. But this means to give up not
only Lorentz-invariance, but also the relativity of simultaneity,
which seems difficult to harmonize with the Michelson-Morley
observations.\\

In conclusion, experiments using series of interferometers should
be of interest to analyze proposals dealing with nonlocality in a
relativistic context. Even those without devices in motion seem
capable of giving us relevant information, at least about how a
theory assuming relativistic nonlocal causality should be
developed, and how the concept of ``wavefunction collapse'' should
not be understood.\\

\section*{Acknowledgements}

I would like to thank Valerio Scarani (EPFL, Lausanne), Nicolas
Gisin, Wolfgang Tittel, Hugo Zbinden (University of Geneva), and an
anonymous Referee for numerous suggestions, and Olivier Costa de
Beauregard (L. de Broglie Foundation, Paris) for stimulating
discussions on retrocausation. It is a pleasure to acknowledge also
discussions regarding experimental realizations with Nicolas Gisin,
Wolfgang Tittel, Hugo Zbinden (University of Geneva), John Rarity,
Paul Tapster (DRA, Malvern), and support by the L\'eman and Odier
Foundations.

\end{document}